\documentclass[aps,prd,showkeys,showpacs,amssymb,%twocolumn,
amsfonts,epsf,preprintnumbers,nofootinbib,superscriptaddress]{revtex4}

\usepackage[dvips]{graphicx}
\usepackage{bm,latexsym,amsmath,amssymb,amsfonts,color}
\def\be {\begin{equation}}
\def\ee {\end{equation}}
\def\ba {\begin{eqnarray}}
\def\ea {\end{eqnarray}}

\def\c  {\gamma}

\def\bi {\begin{itemize}}
\def\ei {\end{itemize}}
\newcommand\beq{\begin{eqnarray}}
\newcommand\eeq{\end{eqnarray}}
\newcommand{\bea}{\begin{eqnarray}}
\newcommand{\eea}{\end{eqnarray}}

\usepackage{graphicx}% Include figure files
\usepackage{dcolumn}% Align table columns on decimal point
\usepackage{bm,morefloats,color}% bold math

\usepackage[
      colorlinks=true,
      linkcolor=blue,
      urlcolor=blue,
      filecolor=blue,
      citecolor=red,
      pdfstartview=FitV,
      pdftitle={},
      pdfauthor={},
      pdfsubject={},
      pdfkeywords={},
      pdfpagemode=None,
      bookmarksopen=true
]{hyperref}

\usepackage{amssymb,bm}

\def\X5sp{{\rm X}_5}
\def\Y3sp{{\rm Y}_3}
\def\Z3sp{{\rm Z}_3}

\begin{document}

\title{Thermodynamics of rotating thin shells in the BTZ spacetime}

\author{Jos\'e P. S. Lemos, Francisco J. Lopes, 
Masato Minamitsuji, and Jorge V. Rocha}
\affiliation{Centro Multidisciplinar de Astrof\'isica - CENTRA,
Departamento de F\'{\i}sica,
Instituto Superior T\'ecnico - IST,
Universidade de Lisboa - UL, Avenida Rovisco Pais 1, 1049-001, Portugal.}

\begin{abstract}
We investigate the thermodynamic equilibrium states of a rotating thin
shell, i.e., a ring, in a (2+1)-dimensional spacetime with a
negative cosmological constant.  The inner and outer regions with
respect to the shell are given by 
the vacuum anti-de Sitter and the rotating
Ba\~{n}ados-Teitelbom-Zanelli spacetimes, respectively.  The
first law of thermodynamics on the thin shell, together with three
equations of state for the pressure, the local inverse temperature and
the thermodynamic angular velocity of the shell, yields the entropy of
the shell, which is shown to depend only on its gravitational radii.
When the shell is pushed to its own gravitational radius and its
temperature is taken to be the Hawking temperature of the
corresponding black hole, the entropy of the shell coincides with the
Bekenstein-Hawking entropy. In addition, we consider simple
ans\"atze for the equations of state, 
as well as a  power-law equation of state where the entropy and the
thermodynamic stability conditions can be examined analytically.
\end{abstract}

%\date{\today}

\maketitle

%%%%%%%%%%%%%%%%%%%%%%%%%
\section{Introduction}

The origin of the Bekenstein-Hawking entropy of a black
hole~\cite{bek1,bch,haw} is one of the greatest mysteries in modern
gravitational physics.  The Bekenstein-Hawking entropy is a measure of
how many Planck areas there are on the event horizon, indicating that
the black hole entropy has its roots in quantum gravity.  However, so
far there has been no fully satisfactory formulation of a quantum
gravity theory and the origin of black hole entropy still remains 
an open question.  An initial development explained the black hole and
its features as the Euclideanized version of its geometry with quantum
gravity making its appearance through the Euclidean functional
integral of the geometry identified essentially as the partition
function for hot gravity \cite{hh}.  Further advancements of the path
integral formalism for black holes, now put in the canonical and grand
canonical ensemble formalism, for Schwarzchild and
Reissner-Nordstr\"om black holes, respectively, were made by York and
followers \cite{york,grand_can1}.  Schwarzchild and
Reissner-Nordstr\"om black holes in 
spacetimes with negative cosmological constant, 
i.e., anti-de Sitter (AdS) spacetimes,
can also have a correlated treatment \cite{hp,grand_can2}.

%%%% Motivation
Since black holes are vacuum solutions of the gravitational field,
while our naive concepts of entropy are based on quantum properties of
matter, it would be useful to study the thermodynamic properties of
collapsing matter, namely whether black hole thermodynamics could
emerge, or not, when we compress matter within its own gravitational
radius. By taking this approach, one expects to obtain indications for
how the Bekenstein-Hawking entropy springs from the final state of
gravitational collapse~\cite{pretisrvol}.  Time-dependent collapsing
solutions are difficult to follow analytically and instead one might
try a sequence of static, or quasistatic, solutions up to the
gravitational radius of a given configuration.  As first shown
in~\cite{quasi_bh}, by using such an approach the black hole
entropy can be recovered.

Therefore it is of great interest to analyze self-gravitating matter
systems, that possess both gravitational and matter degrees of
freedom, and study their thermodynamics.  One of the simplest such
systems is an infinitesimally thin shell where the self-gravitating
matter is confined, placed in an otherwise vacuum
spacetime~\cite{israel}.  The distribution of matter on the shell
fixes the extrinsic curvature, and hence the spacetime geometry
outside the shell, via the junction conditions.  Once the setup is
given, 
one can check under which conditions the thin shell can
be pushed quasistatically to the horizon radius.

%%%%  Approach
The simplest example of a thin shell spacetime assumes staticity and
spherical symmetry. Naturally, the first study on structure and
thermodynamics of thin shells considered the inner spacetime to be
Minkowski and the outer spacetime to be Schwarzschild~\cite{martin}.
By fixing the surface energy density and pressure through the junction
conditions, imposing that the shell has a given local temperature and
using a canonical ensemble, Martinez~\cite{martin} determined the
thermodynamic properties of the shell, characterized by its rest mass
and radius.  However, the gravitational radius, and thus the black
hole limit, was never taken.  This approach draws in many respects
from York's work~\cite{york} where the thermodynamic properties of a
pure Schwarzschild black hole were treated using a canonical ensemble,
i.e., imposing a fixed temperature on some fictitious massless shell
at a definite radius outside the event horizon.

The ensuing nontrivial extension of~\cite{martin} 
was performed for an electrically charged 
shell~\cite{charged}. In that context the inner and 
outer regions are given by the
(3+1)-dimensional Minkowski and Reissner-Nortstr\"om spacetimes,
respectively.  The thermodynamic properties of the charged shell were
characterized by the surface energy density, the pressure, the
electric potential, the temperature and the entropy.  By taking the
shell to its own gravitational radius, and requiring the temperature
of the shell to be given by the black hole Hawking temperature, the
authors of~\cite{charged} found that the entropy reproduced the
Bekenstein-Hawking formula.

Now, these developments appeared in static 
(3+1)-dimensional spacetimes. It is important 
to study other dimensions, lower and higher, 
as well as rotating situations. 
Rotating configurations with shells in 3+1 
dimensions are hopeless, as the Kerr metric
defies being the metric of any reasonable
rotating matter source. 
For small rotation, however, the matching 
problem has solution~\cite{toolkit}
(see also~\cite{muslak}).
In higher dimensions, it is possible 
to find shell solutions in some 
rotating odd-dimensional
spacetimes~\cite{drs,roc}, and thus one could try
to perform a 
thermodynamics analysis in them.
On the other hand, in one lower dimension, 
in 2+1 dimensions, the dynamics and thermodynamics 
of a thin shell is ready for such an incursion.

Indeed, 
in (2+1)-dimensional dimensional spacetimes 
there is the pure black hole solution, the  
Ba\~{n}ados-Teitelboim-Zanelli (BTZ) spacetime~\cite{btz}
that inhabits an AdS background. 
It is a solution that reflects 
in some way the rotating properties of the rotating 
3+1 Kerr solution in an AdS background. 
Moreover, 
the BTZ black hole can be formed via gravitational
collapse of matter shells in a (2+1)-dimensional
spacetime reproducing anew what happens in the 
3+1 world. 
For instance, thin shell, i.e., thin ring, 
collapse
to a BTZ black hole has been studied 
in~\cite{collapse1,collapse2,mop}, 
and spinning string dynamics in a
(2+1)-dimensional AdS background has also been shown to give rise
to a rotating BTZ black hole \cite{kimporrati}.  
Mechanical properties of a stationary shell in a BTZ 
background and its quasistatic collapse up to 
its own gravitational radius were displayed in~\cite{energycond}.
On the other hand,
the termodynamics properties of the 
BTZ black hole have been studied 
thoroughly in~\cite{zaslavskii,bcm,wangwu} (see also 
\cite{btz}).
Thus, in this context, it is of interest 
to study the thermodynamics of self-gravitating thin
shells. 
The approach of~\cite{martin,charged} can be applied to
AdS spacetimes in 2+1 dimensions~\cite{lemos1,lemos2}.
In particular, the authors of 
\cite{lemos2} studied a thin shell
in a static 
BTZ spacetime,
where the inner and outer regions with respect to the shell were given
by the (2+1)-dimensional AdS and BTZ spacetimes,
respectively. The Bekenstein-Hawking entropy formula was
recovered in the limiting case when the shell sat at its own
gravitational radius, provided the shell's temperature coincided with
the temperature of the corresponding BTZ black hole.

Having these previous studies in mind, in the present work we will
investigate the thermodynamic properties of a rotating thin shell
in a (2+1)-dimensional AdS background.  In
the setup we will consider, the interior and exterior of the shell are
described by the static vacuum AdS spacetime in (2+1) dimensions and
the rotating BTZ solution, respectively.  
It is known that the BTZ black
hole also has the Bekenstein-Hawking entropy $S=\frac{A_+}{4G}, $
where $A_+=2\pi r_+$ is the circumference of the outer horizon
(we set the Planck constant, the 
Boltzmann constant, and the velocity 
of light to unity).  Our first goal is to see explicitly
how adding rotation can generalize the thermodynamics on the thin
shell obtained in the static case~\cite{lemos1,lemos2}.  
Our second goal is to compare the thermodynamic
properties of a rotating thin shell to those obtained
in~\cite{charged} for the charged (nonrotating) case. It is well
known that there are similarities between charged and rotating black
holes, e.g., the presence of two gravitational radii, the outer
and inner horizons.  Our analysis
will reveal substantial similarities between charged and rotating thin
shells, irrespective of the dimensionality and asymptotic structure of
the spacetime.

%%%% Outline
This paper is organized as follows.
In Sec.~\ref{sec2}, we study
the mechanical properties of a rotating thin shell in a
(2+1)-dimensional spacetime with a BTZ exterior.  
In Sec.~\ref{sec3}, taking the
locally measured proper mass, the circumference, and the angular
momentum as the independent thermodynamic variables 
on the shell, we consider the
first law of thermodynamics and determine the thermodynamic equations
of state, as well as the entropy.  Sec.~\ref{sec4} is devoted to
the study of the most meaningful equations of state.
This is where we consider the black hole limit for rotating thin
shells and show that the Bekenstein-Hawking entropy is recovered if
the intrinsic temperature of the shell is equal to the Hawking
temperature.  In Sec.~\ref{sec:ChoosingBC} other simple equations
of state for the temperature and the electric potential inspired in
the black hole case are devised.  In Sec.~\ref{sec5}, we study
thermodynamic properties of the rotating thin shell for power-law
equations of state, for which exact expressions for the entropy and
stability conditions are provided.  Finally, we conclude in
Sec.~\ref{sec6}.
In the Appendix we give the BTZ black hole 
thermodynamic properties.

%%%%%%%%%%%%%%%%%%%%%%%%%%%%%%%%%%%%%%%%%%%%%%
\section{Timelike thin shells in the (2+1)-dimensional 
spacetime\label{sec2}}

%%%%%%%%%%%%%%%%%%%%%%%%
\subsection{The outer and inner spacetimes}

We consider Einstein gravity with a cosmological constant in a
(2+1)-dimensional spacetime.  The Einstein equations are given by
(the velocity of light is set to one)
\bea
\label{}
G_{\mu\nu}+\Lambda g_{\mu\nu}=8\pi G T_{\mu\nu},
\eea
where Greek indices $\mu,\nu=0,1,2$ run over time and spatial
coordinates.  Here, $g_{\mu\nu}$ denotes the metric tensor,
$G_{\mu\nu}$ represents its corresponding Einstein tensor, and
$T_{\mu\nu}$ is the energy-momentum tensor for matter in the
(2+1)-dimensional spacetime.  $G$ and $\Lambda$ are the
gravitational and cosmological constants, respectively, and we assume
that $\Lambda<0$, so that the spacetime is asymptotically AdS, with
curvature scale
\bea
\ell=\sqrt{-\frac{1}{\Lambda}}
\label{ell}\,.
\eea

Next we introduce a timelike shell, i.e., a ring in the
(2+1)-dimensional spacetime, with radius $R$, which divides the
spacetime into the outer and inner regions denoted by $M_{(o)}$ and
$M_{(i)}$, respectively.  We also assume that off the shell the
spacetime is vacuum and hence $T_{\mu\nu}=0$ everywhere except at the
location of the shell.  
The spacetime outside the shell ($r>R$) is
described by the rotating BTZ solution, whose metric is given
by~\cite{btz}
\bea
\label{o}
ds_{(o)}^2&=& -\frac{(r^2-r_+^2)(r^2-r_-^2)}
                  {\ell^2 r^2}
               dt_{(o)}^2
???       +\frac{\ell^2 r^2}{(r^2-r_+^2)(r^2-r_-^2)}dr^2
+r^2\Big(d\phi-\frac{r_+ r_-}{\ell r^2}dt_{(o)}\Big)^2\,,
\quad r>R\,,
\eea
where $t_{(o)}$ is the outer time coordinate and $(r,\phi)$ are the
radial and azimuthal coordinates.  The two gravitational radii $r_+$
and $r_-$ are related to the spacetime
Arnowitt-Deser-Misner (ADM) mass $m$ and the angular momentum ${\cal J}$,
respectively, by
\bea
\label{ml}
r_+^2+ r_-^2=
8G\ell^2
m,
\eea 
\bea
r_+r_-
=
4G\ell {\cal J}.
\eea 
We assume  
\bea
r_+\geq r_-\,,
\eea 
which then means
\bea
m\geq \frac{{\cal J}}{\ell}\,.
\eea
The two inequalities are saturated in the extremal case, $r_+=r_-$, 
i.e., $m=\frac{{\cal J}}{\ell}$.
We assume that the shell's character is always timelike
and the shell 
is located outside the event horizon, 
\bea
R>r_+\,.
\eea
Therefore, the
outer region does not contain neither of the horizons $r=r_\pm$ nor
the singularity $r=0$.
The spacetime inside the shell ($r<R$) is pure AdS.
It is the vacuum solution ($m=0$ and ${\cal J}=0$)
of the outer metric, i.e., 
\bea
\label{i}
ds_{(i)}^2= -\frac{r^2} {\ell^2} dt_{(i)}^2
       +\frac{\ell^2}{r^2}dr^2
       +r^2d\phi^2\,,
\quad r<R\,,
\eea
where $t_{(i)}$ is the inner time coordinate, which may differ from
the outer time coordinate $t_{(o)}$.  Concerning the spatial
coordinates $(r,\phi)$, we have not distinguished them from those in
the outer region.  But, as argued below, in order to match these two
metrics smoothly across the timelike hypersurface, the shell must
corotate with the outer spacetime.  Consequently, an angular
coordinate $\psi$ should be introduced instead of $\phi$, such that
exterior spacetime in the new coordinate system $(t_{(o)},r,\psi)$ is
corotating with the shell.
The entire spacetime is vacuum except at the thin shell, i.e., we are
in the presence of a codimension-one distributional source at $r=R$.

The metrics \eqref{o} and \eqref{i} can be collectively expressed by
\bea
\label{collect}
ds_{(I)}^2
&=&-f_{(I)} (r)^2dt_I^2 
+ g_{(I)}(r)^2 dr^2+r^2 \big(d\phi+h_{(I)} (r)dt_{(I)}\big)^2,
\eea
where $I=o/i$ refers to either the outer or inner region with respect
to the shell, and from \eqref{o} and \eqref{i} the functions
$f_{(I)}$, $g_{(I)}$ and $h_{(I)}$ read
\bea
&&f_{(o)}(r)=\frac{\sqrt{(r^2-r_+^2)(r^2-r_-^2)}}
                  {\ell r},
\quad
g_{(o)}(r)=\frac{1}{f_{(o)}(r)},
\quad
h_o(r)=-\frac{r_+r_-}{\ell r^2},
\nonumber\\
&&
f_{(i)}(r)=\frac{r} {\ell},
\quad
g_{(i)}(r)=\frac{1}{f_{(i)}(r)},
\quad
h_{(i)}(r)=0.
\eea

%%%%%%%%%%%%%%%%%%%%%%%%
\subsection{Junction conditions and the shell spacetime}

With the outer and inner regions separated by the shell, the shell
dynamics is determined by the Israel junction
conditions~\cite{israel}.  
The induced metrics on the shell, 
from the outer and inner regions, which we denote generically
by $h^{(I)}_{ab}$,
are given by 
\bea
h^{(I)}_{ab}=e_{(I)a}^\mu e_{(I)b}^\nu g^{(I)}_{\mu\nu}\,, 
\eea
where 
$e_{(I)a}^\mu$ is the
projection tensor to the shell, each viewed from the $(I)$ side.
Then the first junction condition requires
\bea
\label{junc}
\big[h_{ab}\big]=0,
\eea
where $[F]=F_{(o)}-F_{(i)}$ represents the jump of a physical quantity
$F$ across the shell.
Equation~\eqref{junc} ensures the uniqueness of the induced geometry on the
shell $h^{(o)}_{ab}=h^{(i)}_{ab}$. Thus, on the shell
one can define a metic $h_{ab}$ such that
\bea
h_{ab}=h^{(o)}_{ab}=h^{(i)}_{ab}.
\label{uni}
\eea
The second junction condition is given by 
\bea
\label{junc2}
       \big[K_{ab}\big]
     -h_{ab} \big[K\big]
=-8\pi G S_{ab},
\eea
where
\bea 
K^{(I)}_{ab}=e_{(I)a}^\mu e_{(I)b}^\nu 
\nabla^{(I)}_{(\mu} n^{(I)}_{\nu)}
\eea
is the extrinsic curvature tensor defined on the shell's hypersurface,
$\nabla^{(I)}_{\mu}$ is the covariant derivative with respect to the
(2+1)-dimensional metric in region $(I)$ and $n^{(I)}_{\mu}$ is the
unit normal vector viewed from the $(I)$ side.  Moreover, $K=
h^{ab}K_{ab}$ is the trace of the extrinsic curvature, and $S_{ab}$
represents the energy-momentum tensor of matter on the thin shell,
i.e., on the (1+1)-dimensional spacetime intrinsic to the shell.
Equation~\eqref{junc2} determines how the jump to the geometry exterior to
the shell is generated by the matter on the shell.

We now apply these junction conditions to our problem.  As the outer
spacetime is rotating while the inner spacetime is static, in order to
match these two regions, the shell at $r=R$ must corotate with the
outer BTZ region \cite{energycond}
(see also \cite{toolkit}).  
For this purpose, we introduce a coordinate system
corotating with the shell by adopting a new angular coordinate
$d\psi$ such that
\bea
d\psi= d\phi+h_{(I)} (R)dt_{(I)}\,.
\eea
The
line element given in Eq.~\eqref{collect} is then written as
\bea
ds_{(I)}^2
&=&
-f_{(I)} (r)^2dt_{(I)}^2 +g_{(I)}(r)^2 dr^2
+r^2\big(d\psi + {\bar h}_{(I)} (r)dt_{(I)}\big)^2,
\eea
where we have 
introduced 
\bea 
\bar h_{(I)}(r)= h_{(I)} (r)-h_{(I)} (R)\,.
\eea
At
the position of the shell ${\bar h}_{(I)}(R)=0$ and the effects of the
spacetime rotation are hidden at the level of the induced geometry.
The induced line element on the shell, which is uniquely determined
by~\eqref{uni}, is given by
\bea
\label{ind}
ds_\Sigma^2
=h_{ab} dy^a dy^b
=-d\tau^2+R^2d\psi^2,
\quad 
R=R(\tau),
\eea
where Latin indices $a,b=0,1$ run over the (1+1)-dimensional shell's
coordinates, $y^a=(\tau,\psi)$. The proper time on the shell
$\tau$ is defined by
\bea
\label{kint}
d\tau
=\sqrt{ f_{(o)} (R)^2dt_{(o)}^2 -g_{(o)}(R)^2 dR^2}
=\sqrt{ f_{(i)} (R)^2dt_{(i)}^2 -g_{(i)}(R)^2 dR^2},
\eea
which also 
fixes the relation between the outer and inner time coordinates,
$t_{(o)}$ and $t_{(i)}$.  The nonvanishing components of the
projection tensor on the shell, $e_{(I)a}^\mu$, 
viewed from side $(I)$ are given by
\bea
e_{(I)\tau}^t=
\dot{t}_{(I)}=\frac{1}{f_{(I)}(R) }\sqrt{1+g_{(I)}(R)^2\dot{R}^2} ,
\quad
e_{(I)\tau}^r=\dot{R},\quad
e_{(I)\psi}^\psi=1,
\eea
where an overdot represents a derivative with respect to proper time
$\tau$.  The unit normal vector to the shell viewed from side $(I)$,
$n_\mu^{(I)}$, is
\bea
n_\mu^{(I)}
&=&
\Big(
-\dot{R} f_{(I)}(R) g_{(I)}(R)\,,\,
g_{(I)}
\sqrt{1+g_{(I)}(R)^2\dot{R}^2}\,,\,
0\Big)\,,
\eea
and obeys $n^{\mu}_{(I)} n_\mu^{(I)}=1$. 
Since we are interested in the quasistatic process, in the rest of
this paper we assume that $\dot{R}=\ddot{R}=0$.  In this case the
components of the extrinsic curvature tensor are given by
\bea
\label{ext}
K_{(I)}{}^{\tau}{}_{\tau}
&=&\frac{f_{(I)}'(R)} {f_{(I)}(R) g_{(I)}(R)},
\quad
K_{(I)}{}^\psi{}_\psi
=\frac{1}{g_{(I)} (R)R},
\quad
K_{(I)}{}^\tau{}_\psi
=-\frac{ R^2 h_{(I)}'(R)}{2f_{(I)}(R)g_{(I)}(R)},
\eea
where a prime means a derivative with respect to $r$.
For the spacetimes we consider in this work we can use
$f_{(I)}g_{(I)}=1$ to simplify these expressions.  The components of
the extrinsic curvature tensor for the outer region $(o)$ are given by
\bea
K_{(o)}{}^{\tau}{}_\tau
&=&
\frac{R^4-r_+^2r_-^2}   {\ell R^2\sqrt{(R^2-r_+^2)(R^2-r_-^2)}},
\quad
K_{(o)}{}^{\psi}{}_\psi
=\frac{\sqrt{(R^2-r_+^2)(R^2-r_-^2)}}{\ell R^2},
\quad
K_{(o)}{}^\tau{}_\psi
=-\frac{r_+r_-}{\ell R}\,.
\label{ks1}
\eea 
Similarly, those for the inner region $(i)$ are given by 
\bea
K_{(i)}{}^{\tau}{}_\tau
&=&
\frac{1}   {\ell},
\quad
K_{(i)}{}^{\psi}{}_\psi
=\frac{1}{\ell},
\quad
K_{(i)}{}^\tau{}_\psi
=0.
\label{ks2}
\eea 
At this point, we have computed all the quantities necessary to obtain
the energy-momentum tensor of matter on the shell.

%%%%%%%%%%%%%%%%%%%%%%%%
\subsection{The energy-momentum tensor on the shell}

We denote the nonzero components of the energy-momentum tensor of
matter on the shell by
\bea
\label{em}
S^\tau{}_\tau= -\sigma,\quad
S^\psi{}_\psi=p,\quad
S^\tau{}_\psi=j,
\eea
where $\sigma$, $p$ and $j $ represent the energy density, the
pressure, and the angular momentum density of the shell, respectively.
The second junction condition~\eqref{junc2}
plus Eqs.~(\ref{ks1})-(\ref{ks2})
determine the form of
these components
\bea
\sigma
&=&
\frac{1}{8\pi G \ell}
\Big(
1-\frac{1}{R^2}\sqrt{(R^2-r_+^2)(R^2-r_-^2)}
\Big),
\label{sig1}
\\
p
&=&
\frac{1}{8\pi G \ell}
\Bigg(
\frac{R^4-r_+^2 r_-^2}{R^2\sqrt{(R^2-r_+^2)(R^2-r_-^2)}}-1
\Bigg),
\label{p1}
\\
j&=&\frac{r_+r_-}{8\pi G \ell R}\,.
\label{j1}
\eea
In the static, nonrotating, limit,
$j=0$, and so $r_-=0$, we recover 
from Eqs.~(\ref{sig1}) and (\ref{p1})
the result
of~\cite{lemos2},
$\sigma=\frac{1}{8\pi G R}
\left(
\frac{R}{\ell}-\sqrt{\frac{R^2}{\ell^2}-8Gm}
\right)$
and 
$p=\frac{1}{8\pi G}
\frac{R}{\ell^2}
\left(
\frac{1}{\sqrt{\frac{R^2}{\ell^2}-8Gm}}
-\frac{1}{\frac{R}{\ell}}
\right)$,
where we made the replacement $r_+^2\to 8G\ell^2 m$, arising from
Eq.~\eqref{ml}. 
In the presence of generic rotation, the
energy-momentum tensor given by 
Eqs.~(\ref{sig1})-(\ref{j1}) describes an
imperfect fluid, see 
\cite{energycond,mop} (see also \cite{drs}).  
The surface energy density $\sigma$ and pressure $p$
are non-negative, and satisfy $p\geq \sigma$.  Therefore, the matter
shell generically obeys the weak energy condition. However, the
dominant energy condition is violated, except in the extremal
limit $r_+=r_-$, in which case the inequality is
saturated~\cite{energycond}. 

Defining the locally measured
proper mass as $M=2\pi R\sigma$ 
and the angular momentum of the shell as  $J=2\pi R j$
and using Eqs.~(\ref{sig1})-(\ref{j1})
we obtain
\bea
M
&=&
2\pi R\sigma 
=\frac{R}{4G\ell}
\Big(
1-\frac{1}{R^2}\sqrt{(R^2-r_+^2)(R^2-r_-^2)}
\Big),
\label{Mmass1}\\
J
&=&
2\pi R j
= \frac{ r_+r_-}{4G \ell}\,.
\label{JJ1}
\eea
Thus, the angular momentum of the shell
$J$ is independent of the position of the
shell $R$.
From Eq.~(\ref{ml})
we see that it is
identical to that of the exterior 
BTZ spacetime,
\bea
{\cal J}
=J\,.
\eea
This property is very similar to the case of the 
electrically charged
shell~\cite{charged}, where the charge $Q$ does not depend on the
shell position $R$.  
The locally measured proper mass $M$ is related to the
ADM mass $m$ defined in~\eqref{ml} by
\bea
m
=
 \frac{R M}{\ell}
-2GM^2
+\frac{2G}{R^2}J^2\,,
\label{mMJ}
\eea
where the first, second and third terms correspond to the local rest
mass, the gravitational binding energy and the kinetic energy due to
rotation, respectively.

We would like to emphasize that in our case the inner region is pure
$(2+1)$-dimensional AdS spacetime and hence it has locally zero ADM
mass and zero angular momentum.  In the more complex case that the
region inner to the shell contains instead a BTZ black hole then the
total ADM mass and angular momentum of the outer spacetime defined at
infinity would include in addition the ADM mass and angular momentum
of the interior black hole.

%%%%%%%%%%%%%%%%%%%%%%%%%%%%%%%%%%%%%%%%%%%%%%
\section{Thermodynamic entropy of the shell\label{sec3}}

%%%%%%%%%%%%%%%%%%%%%%%%
\subsection{Thermodynamics on the shell}

We now analyze the rotating thin shell system from a thermodynamics
point of view.  We assume that the shell is in thermal equilibrium,
with a locally measured temperature $T$ and entropy $S$.  In the
entropy representation, the entropy $S$ of a system can be expressed
as a function of the state independent variables.  
One can 
take as
state independent thermodynamic 
variables for the thin shell, 
the proper mass $M$, the area of the shell
$A$, 
and the angular momentum $J$. 
Then we can express the entropy as a function of these
quantities, $S=S(M,A,J)$ and the first law of thermodynamics reads
\bea
\label{1st0}
TdS=dM +p\, d A-\Omega\, dJ\,, 
\eea
for $T(M,A,J)$, $p(M,A,J)$, and $\Omega(M,A,J)$, representing 
the temperature, the pressure, and the angular velocity of 
the shell in terms of the state variables.

Now, since in
this (2+1)-dimensional spacetime the area of the shell $A$ is
mathematically equivalent to the position $R$ except for the trivial
factor $2\pi$, we make use of $R$ as the independent variable, instead
of $A$, as it will facilitate the 
calculations.  Then we can express the entropy as a function of these
quantities,
\bea
\label{ent_eos}
S=S(M,R,J)\,.
\eea
Defining further 
the inverse local temperature $\beta$  of the shell as 
\bea
\beta= \frac{1}{T},
\eea
the first law of thermodynamics (\ref{1st0}) now reads
\bea
\label{1st}
dS=\beta\, dM+2\pi p \beta\, dR - \beta\Omega\, dJ.
\eea
The integration of this equation to yield $S=S(M,R,J)$
can then be performed once the equations of state,
\bea
p&=&p(M,R,J)\,,
\label{3eos}\\
\beta&=&\beta(M,R,J)\,,
\label{3eosbeta}\\
\Omega&=& \Omega (M,R,J)\,
\label{3eosomega},
\eea
are specified. 
For $M$, $p$, and $J$  we use 
expressions~\eqref{p1},~\eqref{Mmass1} and~\eqref{JJ1} 
obtained from the junction conditions.  On the
other hand, $\beta$ and $\Omega$ play the role of integration factors,
which must be specified in order to obtain an exact expression for the
entropy.  However, the choice of these functions is constrained by the
necessity to satisfy the integrability conditions that follow directly
from the first law~\eqref{1st},
\bea
\Big(\frac{\partial \beta}{\partial R}\Big)_{J,M}&
=&2\pi\Big(\frac{\partial (\beta p)}{\partial M}\Big)_{J,R},
\label{integra}
\\
2\pi\Big(\frac{\partial (\beta p)}{\partial J}\Big)_{R,M}
&=&-\Big(\frac{\partial (\beta \Omega)}{\partial R}\Big)_{J,M},
\label{integra2}\\
\Big(\frac{\partial \beta}{\partial J}\Big)_{R,M}
&=&-\Big(\frac{\partial (\beta \Omega)}{\partial M}\Big)_{J,R}.
\label{integra3}
\eea

We should also require the local thermodynamic stability conditions to
be satisfied,
\bea
&&
\Big(\frac{\partial^2 S }{\partial M^2}\Big)_{R,J}\leq 0,
\label{sta1}
\\
&&
\Big(\frac{\partial^2 S }{\partial R^2}\Big)_{M,J}\leq 0,
\label{sta2}
\\
&&
\Big(\frac{\partial^2 S }{\partial J^2}\Big)_{M,R}\leq 0,
\label{sta3}
\\
&&
\Big(\frac{\partial^2 S}{\partial M^2}\Big)_{J,R}
\Big(\frac{\partial^2 S}{\partial R^2}\Big)_{M,J}
-\Big\{\Big(\frac{\partial^2 S}{\partial M\partial R}\Big)_{J}\Big\}^2
\geq 0,
\label{sta4}
\\
&&
\Big(\frac{\partial^2 S}{\partial J^2}\Big)_{M,R}
\Big(\frac{\partial^2 S}{\partial R^2}\Big)_{M,J}
-\Big\{\Big(\frac{\partial^2 S}{\partial J\partial R}\Big)_{M}\Big\}^2
\geq 0,
\label{sta5}
\\
&&
\Big(\frac{\partial^2 S}{\partial M^2}\Big)_{J,R}
\Big(\frac{\partial^2 S}{\partial J^2}\Big)_{M,R}
-\Big\{
\Big(\frac{\partial^2 S}{\partial M\partial J}\Big)_{R}\Big\}^2
\geq 0.
\label{sta6}
\eea
For the derivation of the stability conditions with three independent
variables see  Appendix B of~\cite{charged}.

%%%%%%%%%%%%%%%%%%%%%%%%
\subsection{The three equations of state}

\subsubsection{Useful relations}

It is useful to define the redshift function $k$ as 
\bea
&&
k\big(r_+,r_-,R\big)
=\frac{R}{\ell}
\sqrt{
\Big(1-\frac{r_+^2}{R^2}\Big)
\Big(1-\frac{r_-^2}{R^2}\Big)}.
\eea
Then we can 
write~\eqref{Mmass1} 
as
\bea
\label{mass}
M(r_+,r_-,R)=
\frac{1}{4G}
\Big(
\frac{R}{\ell}
-k(r_+,r_-,R)
\Big),
\eea
and repeat ~\eqref{JJ1} 
\bea
\label{JJ2}
J(r_+,r_-)
= \frac{ r_+r_-}{4G \ell}\,,
\eea
so that in thermodynamic terms we see 
that with the use of Eqs.~(\ref{mass}) 
and (\ref{JJ2}), 
we can always change the independent variables 
from $(r_+,r_-,R)$ to $(M,R,J)$ and vice versa.
We now prescribe the three equations of state indicated in~\eqref{3eos}.

\subsubsection{The pressure equation of state}

In this manner, 
we can express the pressure obtained in Eq.~\eqref{p1} as a function 
of $(M,R,J)$,
\bea
\label{press_eq}
p(M,R,J)
=\frac{1}{8\pi G \ell}
\Bigg(
\frac{R^4-r_+^2 (M,R,J)\, r_-^2 (M,R,J)}
{R^3\ell k\big(r_+(M,R,J),r_-(M,R,J),R\big)
}-1
\Bigg)
=\frac{MR^3-4 G\ell J^2}
       {2\pi R^3 (R-4G\ell M)},
\eea
which determines the pressure equation of state of the shell.  We find
it more convenient to work with the equations expressed as
functions of $(r_+,r_-,R)$ but we always keep in mind that they depend
implicitly on $M$ and $J$ through $r_\pm (M,R,J)$.

\subsubsection{The temperature equation of state}

Next we turn to the other equation of state, $\beta=\beta (M,R,J)$ 
which is constrained by the integrability condition~\eqref{integra}.
First note that
\bea
\label{pres}
p=-\frac{1}{2\pi}\Big(\frac{\partial M}{\partial R}\Big)_{r_+,r_-}\,,
\eea
which expresses the conservation of the shell's stress-energy tensor.
Then
$
\frac{1}{\beta}\Big(\frac{\partial \beta}{\partial R}\Big)_{r_+,r_-}
=
\frac{1}{\beta}
\Big\{
\Big(\frac{\partial \beta}{\partial R}\Big)_{M,J}
+
\Big(\frac{\partial M}{\partial R}\Big)_{r_+,r_-}
\Big(\frac{\partial\beta}{\partial M}\Big)_{R,J}
\Big\}
=
\frac{1}{\beta}
\Big\{
2\pi\beta
\Big(\frac{\partial p}{\partial M}\Big)_{R,J}
+
\Big[
2\pi p
+\Big(\frac{\partial M}{\partial R}\Big)_{r_+,r_-}
\Big]
\Big(\frac{\partial\beta}{\partial M}\Big)_{R,J}
\Big\}  
=2\pi \Big(\frac{\partial p}{\partial M}\Big)_{J,R}
=\frac{R^4-r_+^2r_-^2}{R(R^2-r_+^2)(R^2-r_-^2)}
=\frac{1}{k}\left(\frac{\partial k}{\partial R}\right)_{r_+,r_-}
$,
where we used Eq.~\eqref{pres}.
Thus, in brief
\bea
\label{temp_eq}
\frac{1}{\beta}\Big(\frac{\partial \beta}{\partial R}\Big)_{r_+,r_-}
=\frac{1}{k}\left(\frac{\partial k}{\partial R}\right)_{r_+,r_-}\,.
\eea
By integrating Eq.~\eqref{temp_eq}, we obtain the inverse 
local temperature equation of state of the shell
\bea
\label{temp}
\beta(r_+,r_-,R)=
k (r_+,r_-,R) b(r_+,r_-),
\eea
where $b(r_+,r_-)$ is an arbitrary function of $r_\pm$.
The function $b(r_+,r_-)$
is interpreted as the temperature of the shell located at
the radius $
R
=\Big\{\frac{1}{2}
\Big(
\ell^2 +r_+^2+r_-^2
+\sqrt{\ell^4 + (r_+^2-r_-^2)^2+2\ell^2 (r_+^2+r_-^2)}
\Big)
\Big\}^{\frac{1}{2}}
$.
Since $k=\sqrt{-g^{(o)}_{tt}(R)}$, the formula \eqref{temp} expresses the
gravitational redshift of the temperature of the shell and is nothing
but the Tolman relation for the temperature in the gravitational
system.  The function $b(r_+,r_-)$ depends on $(M,R,J)$ only through
$r_\pm$.  The integrability condition does not yield a precise form
for $b(r_+,r_-)$, which depends on the properties of matter
within the
shell.

\subsubsection{The angular velocity equation of state}

Next, we consider $\Omega=\Omega(M,R,J)$.  Using the integrability
conditions~\eqref{integra2} and (\ref{integra3}), 
together with relation~\eqref{pres}, we find
$
\Big( \frac{\partial p}{\partial J}\Big)_{M,R}
+\Omega \Big( \frac{\partial p}{\partial M}\Big)_{J,R}
=
p\Big(\frac{\partial \Omega}{\partial M}\Big)_{J,R}
-\frac{1}{2\pi}\Big(\frac{\partial \Omega}{\partial R}\Big)_{M,J}
=-\frac{1}{2\pi}
\Big\{
\Big(\frac{\partial \Omega}{\partial R}\Big)_{M,J}
+
\Big(\frac{\partial M}{\partial R}\Big)_{r_+,r_-}
\Big(\frac{\partial \Omega}{\partial M}\Big)_{J,R}
\Big\}
=-\frac{1}{2\pi}
\Big(\frac{\partial \Omega}{\partial R}\Big)_{r_+,r_-}
$.
Then, taking in consideration Eq.~\eqref{temp_eq}, 
$\Omega$ obeys
$\Big(\frac{\partial (\Omega \beta)}{\partial R}\Big)_{r_+,r_-}
=-2\pi \beta 
\Big(\frac{\partial p}{\partial J}\Big)_{M,R}
=\frac{2r_+r_-b(r_+,r_-)}{\ell R^3}$, 
where we used Eq.~\eqref{temp} in the last step.
So, in brief
\bea
\label{angle_eq}
\Big(\frac{\partial (\Omega \beta)}{\partial R}\Big)_{r_+,r_-}
=\frac{2r_+r_-b(r_+,r_-)}{\ell R^3}\,.
\eea
After integrating Eq.~\eqref{angle_eq} we obtain 
the angular velocity equation of state
\bea
\label{omega}
\Omega(r_+,r_-,R)=\frac{ r_+r_-}{\ell k(r_+,r_-,R)}
\Big(
c(r_+,r_-)-\frac{1}{R^2}
\Big),
\eea
where $c(r_+,r_-)$ is an arbitrary function of $r_\pm$.

\subsubsection{In a nutshell}

In this way, we have found the three equations of
state~\eqref{press_eq}, \eqref{temp} and~\eqref{omega}, which are
necessary to determine the entropy of the shell, as we will argue 
below.  Up to now, the integration of the integrability
conditions has introduced two integration constants, $b(r_+,r_-)$ and
$c(r_+,r_-)$, which are free functions of $r_+$ and $r_-$.

%%%%%%%%%%%%%%%%%%%%%%%%
\subsection{The entropy of the shell}

By changing variables from $(M,R,J)$ to $(r_+,r_-,R)$ and substituting
Eqs.~\eqref{mass}, \eqref{press_eq}, \eqref{temp} and~\eqref{omega}
into the first law of thermodynamics, Eq.~\eqref{1st}, we obtain
\bea
\label{ent}
dS
&=&
\frac{b}{8 G \ell^2}
\Big[
\big(1-r_-^2c(r_+,r_-) \big)dr_+^2
+
\big(1-r_+^2c(r_+,r_-) \big)dr_-^2
\Big].
\eea
This expression implies that the two integration constants must
satisfy the integrability condition
\bea
\label{integ2}
\frac{\partial b}{\partial r_-^2}
\Big(
1-r_-^2 c
\Big)
-br_-^2\frac{\partial c}{\partial r_-^2}
=
\frac{\partial b}{\partial r_+^2}
\Big(
1-r_+^2 c
\Big)
-br _+^2\frac{\partial c}{\partial r_+^2}.
\eea
This condition can be equivalently expressed as
\bea
\frac{\partial b}{\partial r_+^2} - \frac{\partial b}{\partial r_-^2}
= \frac{\partial(b c)}{\partial \log(r_+^2)} - \frac{\partial(b
c)}{\partial \log(r_-^2)},
\eea
which makes it manifest that any choice for which $b$ is a function of
$r_+^2+r_-^2$ only, and $bc$ is a function of $r_+^2 r_-^2$ only, will
satisfy the integrability condition. In other words,
Eq.~\eqref{integ2} is automatically obeyed whenever $b$ and $bc$ are
only functions of the ADM mass $m$ and the angular momentum $J$,
respectively. This will be used below when we search for equations of
state.
However, in generic cases, in order to obtain a specific 
expression for the entropy we
need to choose either $b(r_+,r_-)$ or $c(r_+,r_-)$, and then obtain
the remaining function by integrating Eq.~\eqref{integ2}.

Relation~\eqref{ent} also indicates that the entropy $S$ is a function
of $r_+$ and $r_-$ only,
\bea
S=S(r_+,r_-),
\eea
and hence a function of $(M,R,J)$ only through $r_\pm(M.R,J)$,
\bea
\label{implicit}
S(M,R,J)
= S\big(r_+ (M,R,J),r_-(M,R,J)\big). 
\eea 
It is also worth mentioning that, from \eqref{implicit}, shells with
the same $r_+$ and $r_-$, namely with the same ADM mass $m$ and
angular momentum $J$ but at a different position $R$, have the same
entropy.  Thus, an observer measuring $m$ and $J$ cannot distinguish
shells with different radii by measuring the entropy.

%%%%%%%%%%%%%%%%%%%%%%%%%%%%%%%%%%%%%%%%%%%%%%
\section{The thin shell and the black hole limit\label{sec4}}

%%%%%%%%%%%%%%%%%%%%%%%%
\subsection{A precisely chosen temperature equation 
of state and the entropy}

As the equation of state for the inverse temperature
$b(r_+,r_-)$, let us take it
to be of the form
\bea
\label{b_form}
b(r_+,r_-) = b_+ \gamma, %= 2\pi\ell^2\frac{r_+}{r_+^2-r_-^2}\gamma,
\eea
where $b_+$ is the inverse Hawking temperature of the BTZ black
hole given by 
\bea
\label{hawkingt}
b_+= 2\pi\ell^2\frac{r_+}{r_+^2-r_-^2}\,
\eea
and
$\gamma$ is a parameter which will depend on the properties of matter
on the shell.  We assume $\gamma>0$, so that the temperature is
positive.  This is one of the simplest possible temperature equations of
state, setting the shell's fluid temperature proportional to the black
hole temperature.

We also have to specify $c(r_+,r_-)$, so that it satisfies the
integrability condition \eqref{integ2}.  There is a family of
solutions for $c$, but here we choose the following
particular solution
\bea
\label{c_form}
c(r_+,r_-)=\frac{1}{r_+^2},
\eea 
which makes the angular velocity $\Omega$ vanish when $R\to r_+$ [see
Eq.~\eqref{omega}].  By substituting \eqref{b_form} and \eqref{c_form}
into Eq.~\eqref{ent}, we obtain the differential for the entropy of
the shell
\bea
\label{diff_p}
dS&=&\frac{\gamma}{4G}dA_+,
\eea
where $A_+=2\pi r_+$ represents the circumference (area) of the event
horizon.  By integrating \eqref{diff_p}, the entropy of the shell is
given by
$
S=S_0+\frac{\gamma}{4G}A_+
$,
where $S_0$ is an integration constant.  Requiring that when the shell
is absent --- or equivalently when $M=0$ and $J=0$ ($r_+=r_-=0$ from
\eqref{Mmass1}) --- the entropy vanishes, we fix $S_0=0$ and obtain
\bea
\label{ent_integ}
S=\frac{\gamma}{4G}A_+,
\eea
which shows that the entropy of the rotating shell depends on
$(M,R,J)$ only through $r_+$.  We note that for the entropy
\eqref{ent_integ} all the thermodynamic stability conditions
\eqref{sta1}-\eqref{sta6} are satisfied provided $\gamma>0$.  The
parameter $\gamma$ should be determined by the properties of matter on
the shell and cannot be determined a priori.

%%%%%%%%%%%%%%%%%%%%%%%%%%%%55
\subsection{The black hole limit}

Although $\gamma$ should be determined by the properties of matter on
the shell, there is a case in which the properties of the shell have
to be adjusted to the environment. 
Such a situation occurs when the shell is pushed to its own
gravitational radius, $R\to r_+$.  In fact, as the shell approaches
its gravitational radius, quantum effects would be inevitably present
and their backreaction would invalidate the classical treatment we
have adopted, unless we choose the black hole Hawking 
temperature for the
temperature of the shell.  Therefore, we must choose $\gamma=1$, or
equivalently $b=b_+$, see Eq.~(\ref{hawkingt}), 
with $c_+$ still being given by 
Eq.~(\ref{c_form}).  In this case, \eqref{ent_integ} becomes
\bea
\label{bh}
S=\frac{A_+}{4G},
\eea
which is the same as the Bekenstein-Hawking entropy
for the corresponding black hole
(see the Appendix).
Thus, when we push the shell to its gravitational radius the entropy
coincides with the Bekenstein-Hawking entropy.  In the limit $R\to
r_+$, the pressure given by~\eqref{press_eq} diverges as $\frac{1}{k}$
(assuming the spacetime is not extremal, $r_+\neq r_-$), whereas the
angular velocity expressed in~\eqref{omega} vanishes, at least for the
particular choice of the function $c$ made in Eq.~\eqref{c_form}.
Nevertheless, the local inverse temperature \eqref{temp} is
proportional to $k$, so the local temperature of the shell also
diverges as $\frac{1}{k}$.  These divergences cancel out precisely, so
that they can reproduce the Bekenstein-Hawking entropy.
In fact, the first law~\eqref{1st}, 
$dS=\beta\, dM+2\pi p \beta\, dR - \beta\Omega\, dJ
$,
reveals that in the black hole limit, the only term that survives (and
remains finite) in the right-hand side is the pressure term, which
neatly combines with the inverse temperature to yield an area law for
the entropy~\cite{quasi_bh}.

Our approach adds in a nontrivial manner to the results  
of the  static (2+1)-dimensional 
studies presented in~\cite{lemos1,lemos2}, 
having also affinities to the works \cite{martin,charged}. 
This manner of calculating the entropy of a black hole
shares also certain
similarities with the work~\cite{pretisrvol}, in the sense that both studies
consider matter distributed on thin shells to determine the entropy of
black holes.  Here, we used a radially static thin shell that
decreases its radius adiabatically toward its gravitational radius,
maintaining quasistaticity of the spacetime.  On the other hand,
\cite{pretisrvol} considered a reversible contraction of a thin
shell and found that the black hole entropy can be defined as the
thermodynamic entropy stored in matter compressed into a thin layer at
its own gravitational radius.

We also note that the extremal limit $r_+=r_-$ is well defined.  In
this limit, we find that the temperature 
$\frac{1}{b(r_+,r_-)}\to 0$, but the entropy of
the extremal black hole is still given by~\eqref{bh}.  It is
well known that the entropy of an extremal black hole requires special
care.  If we had started our analysis directly using the metric for
the extremal black hole, we would have found a more complicated
expression for the entropy.

%%%%%%%%%%%%%%%%%%%%%%%%
\section{Other equations of state with $b(r_+,r_-)$ and $c(r_+,r_-)$ 
of black hole type \label{sec:ChoosingBC}}

In the previous subsections we imposed a temperature equation of state
of the Hawking type~\eqref{b_form}, as well as the specific
thermodynamic angular velocity~\eqref{c_form}, and obtained that the
entropy of the shell is proportional to $A_+$.  Moreover, if we set
the temperature of the shell exactly equal to the Hawking temperature
(see Eq.~\eqref{hawkingt}, or Eq.~\eqref{b_form} with $\gamma=1$),
then the entropy of the shell precisely reproduces the
Bekenstein-Hawking entropy.  The choices made for the equations of
state~\eqref{b_form} and~\eqref{c_form}, although mandatory for a
shell reproducing the Bekenstein-Hawking area law when approaching its
gravitational radius $r_+$, are just the simplest ones among a larger
class of equations of state allowed by the integrability
condition~\eqref{integ2}.

%%%%%%%%%%%%%%%%%%%%%%%%%%%%%%%%%%%%%%%%%%%%%%%%%%%%
%\subsection{Other four equations of state}

Here, we briefly consider some other choices for the equations of state.

\paragraph*{Case 1.}
For the temperature equation of state~\eqref{b_form}, the
integrability condition~\eqref{integ2} gives a general equation
of state for the thermodynamic angular velocity
\bea
\label{c_general}
c(r_+,r_-)=\frac{1}{r_+^2}\Big(1+ r_+ (r_+^2-r_-^2)
\tilde{c}(r_+^2r_-^2)\Big),
\eea 
where $\tilde{c}(r_+^2r_-^2)$ is an arbitrary function of 
the product $r_+^2 r_-^2$.
Substituting Eqs.~\eqref{b_form} and~\eqref{c_general} into
\eqref{ent} and integrating, we obtain
\bea
S(r_+,r_-)=\frac{\gamma}{4G}\Big(A_+ -\pi  
\int_0^{r_+^2 r_-^2}dx \, \tilde{c}(x)\Big),
\eea
where we set the integration constant to zero, so that $S$ vanishes
when $r_+\to 0$.  If we set $\tilde{c}(r_+^2r_-^2)=0$, we recover
expression~\eqref{ent_integ} which reproduces the Bekenstein-Hawking
entropy for $\gamma=1$.

\paragraph*{Case 2.}
On the other hand, if we start from the angular velocity equation of
state~\eqref{c_form}, then the integrability condition~\eqref{integ2}
allows the quite general equation of state for the inverse temperature
given by 
\bea
\label{b_general}
b(r_+,r_-)=\frac{2\pi\ell^2 h_+(r_+^2)}{r_+^2-r_-^2},
\eea
where $h_+(r_+^2)$ is an arbitrary function of $r_+^2$.
Substituting~\eqref{c_form} and~\eqref{b_general} into~\eqref{ent} and
integrating, we obtain
\bea
S(r_+)= \frac{\pi}{4G}\int_0^{r_+^2}dx \frac{h_+(x)}{x},
\eea
where once again we set the integration constant to zero.  If we set
$h_+(r_+^2)=\sqrt{r_+^2}$, we recover \eqref{ent_integ} which
reproduces the Bekenstein-Hawking entropy for $\gamma=1$.

\paragraph*{Case 3.}
Similarly, if we start from the angular velocity equation of  state
\bea
\label{c-}
c(r_+,r_-)=\frac{1}{r_-^2},
\eea
then the integrability condition~\eqref{integ2} allows the more
general equation of state for the inverse temperature given by
\bea
\label{b_general2}
b(r_+,r_-)=\frac{2\pi\ell^2 h_-(r_-^2)}{r_-^2-r_+^2},
\eea
where $h_-(r_-^2)$ is an arbitrary function of $r_-^2$.
Substituting~\eqref{c-} and~\eqref{b_general2} into~\eqref{ent} 
and integrating, we obtain 
\bea
S(r_-)= \frac{\pi}{4G}\int_0^{r_-^2}dx \frac{h_-(x)}{x}. 
\eea

\paragraph*{Case 4.}
Finally,  if we start from the angular velocity equation of  state,
\bea
\label{c0}
c(r_+,r_-)={\tilde c}(r_+^2 r_-^2),
\eea
where ${\tilde c}(r_+^2r_-^2)$ is an arbitrary function of 
the product 
$r_+^2 r_-^2$, then the integrability condition~\eqref{integ2} allows 
\bea
\label{b0}
b(r_+,r_-)=b_0\ell^2,
\eea
where $b_0$ is an arbitrary constant.
Substituting~\eqref{c0} and~\eqref{b0} into~\eqref{ent} and 
integrating, we obtain 
\bea
S(r_+,r_-)= \frac{b_0}{8G}
\Big(
  r_+^2+ r_-^2
-\int_0^{r_+^2 r_-^2} dx \,{\tilde c}(x)
\Big). 
\eea

The above four cases are the counterparts of those considered for the
charged shell in \cite{charged}.  As in \cite{charged}, we will not
explore them further.

%%%%%%%%%%%%%%%%%%%%%%%%%%%%%%%%%%%%%%%%%%
\section{Power-law equations of state\label{sec5}}

%%%%%%%%%%%%%%%%%%%%%%%%
\subsection{The entropy}

In this section we will focus on
power-law equations of state for both the inverse temperature and the
thermodynamic angular velocity, for which the stability conditions can
be studied explicitly.
We start by specifying the temperature equation of state $b(r_+,r_-)$
in Eq.~\eqref{temp}.  One of the simple but reasonable choices for the
temperature is a power-law function of $r_+^2+r_-^2$ which is related
to the ADM mass $m$ via Eq.~\eqref{ml},
\bea
\label{power_b}
b(r_+,r_-)=
4 G \ell^2 a
\big(r_+^2+r_-^2\big)^{\frac{\alpha}{2}},
\eea
where $a$ and $\alpha$ are free parameters which reflect the
properties of matter on the shell.
For such a temperature equation of state, the integrability
condition~\eqref{integ2} admits the following general solution for the
equation of state for the thermodynamic angular velocity $c(r_+,r_-)$
in Eq.~\eqref{omega},
\bea
c(r_+,r_-)=\frac{ {\tilde c}
(r_+^2r_-^2)}{(r_+^2+r_-^2)^\frac{\alpha}{2}},
\eea 
where ${\tilde c}(r_+^2r_-^2)$ is an arbitrary function of the product
$r_+^2 r_-^2$.  Since the product $r_+^2r_-^2$ is related to the
angular momentum $J$ in Eq.~\eqref{Mmass1}, it is also reasonable that
we introduce another power-law form for ${\tilde c}(r_+^2r_-^2)$, such as
\bea
\label{power_c}
c(r_+,r_-)=\frac{\sigma\,(r_+^2r_-^2)^{\frac{\delta}{2}} }
                        {(r_+^2+r_-^2)^{\frac{\alpha}{2}}},
\eea
where $\sigma$ and $\delta$ are further free parameters which also reflect
the properties of matter on the shell.
Substituting~\eqref{power_b} and~\eqref{power_c} into~\eqref{ent} and
integrating, we obtain
\bea
S(r_+,r_-)
=a
\Big(
\frac{(r_+^2+r_-^2)^{\frac{\alpha}{2}+1}}
       {\alpha+2}
-\frac{\sigma(r_+^2r_-^2)^{\frac{\delta}{2}+1}}{\delta+2}
\Big),
\eea 
where we have set the integration constant $S_0$ to zero, so that the
entropy $S$ vanishes in the limit of $m\to 0$ and $J\to 0$, namely
$r_+\to 0$ and $r_-\to 0$.  This requirement can be satisfied if we
impose the following conditions on the exponents,
\bea
\label{imp2}
\alpha>-2, \quad \delta>-2.
\eea
As in the cases discussed in the previous sections, the entropy of the
rotating shell depends on $(M,R,J)$ only through $r_\pm(M,R,J)$.  We
impose positivity of the temperature, which gives the following
constraint on the parameter $a$,
\bea
\label{imp}
a>0.
\eea

%%%%%%%%%%%%%%%%%%%%%%%%
\subsection{Thermodynamic stability conditions}

We now address the thermodynamic stability of the system.  First, we
consider the stability conditions that do not involve $\sigma$ and
$\delta$. Defining
\bea
\xi_\pm=\frac{r_\pm}{R}\,,
\eea
condition~\eqref{sta1} gives
\bea
\label{cond1}
\alpha
\leq 
\frac{\xi_+^2+\xi_-^2}{\kappa^2}.
\eea
where we have also defined 
\bea
\kappa (\xi_+,\xi_-)
=\sqrt{
\big(1-\xi_+^2\big)
\big(1-\xi_-^2\big)
}\,.
\eea
Note that $0\leq\kappa<1$, with the first inequality being saturated
only when the shell is taken to its gravitational radius, $R\to r_+$.
Condition \eqref{sta2} gives
\bea
\label{cond2}
\alpha
\leq -3\xi_+^2\xi_-^2
\frac{\xi_+^2+\xi_-^2}
       {\big(1-\kappa(\xi_+,\xi_-)-
     \xi_+^2\xi_-^2\big)^2}
\leq 0.
\eea
Finally, condition \eqref{sta4} yields 
\bea
\label{cond3}
\alpha
\leq 
\alpha_\ast(\xi_+,\xi_-)
=
-
\frac{
\big(1+
3\xi_+^2\xi_-^2\big)
\big(\xi_+^2+\xi_-^2\big)}
{\big(1-
\xi_+^2\xi_-^2\big)^2
-\big(
 1+
3\xi_+^2\xi_-^2
 \big)
\kappa^2(\xi_+,\xi_-)}.
\eea
We find that among the above conditions on $\alpha$, condition
\eqref{cond3} is the most stringent.  Noting that
$\alpha_\ast(\xi_+,\xi_-)\leq \alpha_\ast(\xi_-,\xi_-)$ [assuming
$\alpha_\ast(\xi_-,\xi_-)>-2$], where
\bea
\alpha_\ast (\xi_-,\xi_-)
=-\frac{1+3\xi_-^4}
           {(1-\xi_-^2)^3}<-1,
\eea
we conclude that there can be a parameter region where
condition~\eqref{imp2} is met.  The corresponding value of $\xi_-$
is given by
\bea
\xi_-<\sqrt{\frac{1}{2}(1+3^{\frac{1}{3}}-3^{\frac{2}{3}})}=0.4255.
\eea
Finally, we turn to the thermodynamic stability conditions which 
involve $\sigma$ and $\delta$.
Defining
\bea
\tilde{\sigma}=\sigma\, (1+\delta) R^{2(1+\delta)-\alpha},
\eea
condition~\eqref{sta3} gives
\bea
\label{cond4}
-(\xi_+^2+\xi_-^2)^{\frac{\alpha}{2}}
\Big(\xi_+^2+\xi_-^2+\alpha
\xi_+^2\xi_-^2\Big)
+\tilde{\sigma} (\xi_+^2 \xi_-^2)^{\frac{\delta}{2}} 
    \Big(\xi_+^2+\xi_-^2\Big)
\geq 
0.
\eea
Condition~\eqref{sta5} gives
\bea
\label{cond5}
&&-\big(\xi_+^2+\xi_-^2\big)^{\frac{\alpha}{2}}
\Big[
\xi_+^2\xi_-^2
\Big(\xi_+^2+\xi_-^2\Big)
-\alpha (1-\kappa(\xi_+,\xi_-))
\Big(1-\kappa(\xi_+,\xi_-) +
2\xi_+^2\xi_-^2\Big)
\Big]
\nonumber\\
&-&
\tilde{\sigma}
(\xi_+^2 \xi_-^2)^{\frac{\delta}{2}}
\Big[
3\xi_+^2\xi_-^2
\Big(\xi_+^2+\xi_-^2\Big)
+\alpha
\Big(1-\kappa(\xi_+,\xi_-) -
\xi_+^2\xi_-^2\Big)^2
\Big]
\geq 0.
\eea
Finally, condition~\eqref{sta6} gives
\bea
\label{cond6}
&&-\big(\xi_+^2+\xi_-^2\big)^{\frac{\alpha}{2}}
\Big[
\xi_+^2+\xi_-^2
-\alpha
\Big(
\kappa^2(\xi_+,\xi_-)
-
\xi_+^2\xi_-^2
\Big)
\Big]
+\tilde{\sigma}
(\xi_+^2 \xi_-^2)^{\frac{\delta}{2}}
\Big(
\xi_+^2+\xi_-^2
-\alpha \kappa^2(\xi_+,\xi_-)
\Big)
\geq 0.
\eea
In the black hole limit, $\xi_+\to1$, this condition is equivalent to
Eq.~\eqref{cond4}.

If one assumes that the parameter $\sigma$ is dimensionless, so as to
not introduce any further length scales in the problem, inspection of
any of the above conditions, or of Eq.~\eqref{power_c}, shows that the
exponents $\alpha$ and $\delta$ must be related through
\bea
\label{tune}
\delta=\frac{\alpha}{2}-1,
\eea
for dimensional consistency.
This reduces the number of relevant dimensionless parameters down to
four, namely $(\alpha, \tilde{\sigma}, \xi_+, \xi_-)$.

The stability conditions \eqref{cond4}, \eqref{cond5} and
\eqref{cond6} are somewhat complicated. 
 %In order to investigate
%whether they can be compatible with each other and with~\eqref{cond3}.
%With these choices our parameter space is reduced to a
%four-dimensional space, namely $(\alpha, \tilde{\sigma}, \xi_+, \xi_-)$. 
We find numerically
that all the thermodynamic stability requirements [\eqref{imp2},
\eqref{imp}, \eqref{cond3}, (\ref{cond4}-\ref{cond6})] can be met for
a certain region of the parameter space, as long as
\bea
\tilde{\sigma}>0.
\eea
If that is the case, then all stability conditions tend to be
satisfied for a small $\xi_-$, i.e., for slowly rotating shells.  
Thermodynamic stability can also be
insured for shells close to the black hole limit, $\xi_+\to1$, but
this happens only if $\xi_-$ remains small, i.e., far from
extremality.  Moreover, we observe that these conditions are weakly
dependent on $\alpha$.

%%%%%%%%%%%%%%%%%%%%%%%%%%%%%%%%%%%%%%5
\section{Conclusions\label{sec6}}

In this work we have 
investigated the thermodynamic properties of a
rotating thin shell in a (2+1)-dimensional asymptotically AdS
spacetime, where the interior and exterior of the shell were taken to
be the vacuum AdS spacetime
and the BTZ black hole spacetime, respectively.  These two
geometries were matched across the shell using the Israel junction
conditions, which in turn provide the three quantities which
characterize the properties of the matter on the shell, namely, the
surface energy density $\sigma$, the pressure $p$, and the angular
momentum density $j$.  Multiplying the surface energy density and the
angular momentum density by the circumference $2\pi R$, we obtained
the locally measured proper mass $M=2\pi R \sigma$ and the angular
momentum $J= 2\pi R j$ of the shell, thus completing the study of its
mechanical properties.

To address the thermodynamics of the system we adopted the locally
measured mass $M$, the volume of the shell $2\pi R$ and the angular
momentum $J$ as the state independent variables.  Using the first law
of thermodynamics and the equations of state, we were able to obtain
the entropy of the shell $S=S(M,R,J)$.  Due to the additional variable
$J$, the construction of the thermodynamics is more complicated than
in the nonrotating case~\cite{lemos2}.  In order to obtain an
expression for the entropy of the shell, one must specify equations of
state for the pressure $p=p(M,R,J)$, the local inverse temperature
$\beta=\beta (M,R,J)$ and the thermodynamic angular velocity
$\Omega=\Omega(M,R,J)$.  While the pressure equation of state is
automatically determined by the junction conditions, there is a
certain freedom to choose the equations of state for the local inverse
temperature and thermodynamic angular velocity, which only have to
satisfy an integrability condition derived from the first law of
thermodynamics.

One of our main results is that the entropy of the shell must be a
function of the two gravitational radii $r_\pm$ alone.  In particular,
shells with the same $r_\pm$, and thus 
the same ADM mass $m$ and
angular momentum $J$, but at different radii $R$ share the same
entropy.  Thus, from the entropic properties only, it is not
possible to distinguish a shell near the gravitational radius $r_+$
from one asymptotically far.  These findings corroborate
the results obtained recently for charged thin shells in (3+1)
dimensions~\cite{charged}, with the analogy becoming manifest if one
replaces ``electric charge'' and ``Coulomb potential'' with ``angular
momentum'' and ``angular velocity'', respectively.

The integrability conditions allow a multitude of equations of state
for the local inverse temperature $\beta(M,R,J)$ and angular velocity
$\Omega(M,R,J)$.  Choosing a well-motivated temperature equation of
state, namely setting the shell's temperature equal to the BTZ
black hole Hawking temperature, and the simplest possible angular
velocity equation of state that is consistent with that choice, the
resulting entropy of the shell precisely agrees with the
Bekenstein-Hawking entropy of the BTZ black hole.

Nevertheless, many other equations of state that do not yield a simple
area law for the entropy are also consistent with the integrability
condition.  We have presented a large class of such examples, for
which the inverse temperature and angular velocity equations of state
are described by power laws.  For this family of solutions the number
of parameters characterizing the system is quite large.  Excluding the
AdS scale $\ell$ and the radial location of the shell $R$, we are
still left with six parameters.  Through a thermodynamic stability
analysis we have obtained several constraints on the equations of
state parameters.

There remains the possibility of the existence of further classes of
consistent choices for the equations of state, which should require a
dedicated study of their thermodynamic stability properties.

%%%%%%%%%%%%%%%%%%%%%%%%%%%%%%%
\section*{ACKNOWLEDGEMENT}
We thank FCT-Portugal for financial support through
Project No.~PEst-OE/FIS/UI0099/2014.
M.M. was supported by FCT-Portugal through Grant
No.~SFRH/BPD/88299/2012.  J.V.R. acknowledges financial support
provided under the European Union's FP7 ERC Starting Grant ``The
dynamics of black holes: testing the limits of Einstein's theory''
grant agreement No.~DyBHo256667.

\appendix*

\section{\uppercase{Thermodynamics of the BTZ black hole}}
\label{app}

Here we review the thermodynamic
properties of the rotating BTZ black hole
following the Hawking-Page formalism for 
the (3+1)-dimensional Schwarzschild AdS black 
hole \cite{hp}, see 
also \cite{zaslavskii,bcm,wangwu}
who applied diverse formalisms to the same BTZ black hole.  
The Euclidean action $S_E$ in the
(2+1)-dimensional Einstein gravity is given by
\bea
\label{act0}
S_E= -\frac{1}{16\pi G}\int d^3 x\sqrt{g_E}
      \Big(R_E-2\Lambda\Big)\,,
\eea
with $\sqrt{-\Lambda}= \frac{1}{\ell}$ and 
where $R_E$ is the Euclidean Ricci scalar 
given by $R_E=-\frac{6}{\ell^2}$.
The vacuum BTZ Euclidean metric is given by metric 
(\ref{o}) with the time Euclideanized, i.e., $t=i\tau$,
\bea
\label{vaceucbtz}
ds^2= f^2(r)
               d\tau^2
      +\frac{1}{f^2(r)}dr^2
+r^2\Big(d\phi-\frac{ir_+ r_-}{\ell r^2}d\tau\Big)^2\,,
\quad 0<r<\infty\,,
\eea
where 
\bea
\label{fr}
f(r)=\frac{\sqrt{ (r^2-r_+^2)(r^2-r_-^2)}}
              {\ell r}\,.
\eea
The two radii $r_+$
and $r_-$ are now the event horizon and the 
Cauchy horizon radii, respectively, still
related to the spacetime
Arnowitt-Deser-Misner (ADM) mass $m$ and the angular momentum ${\cal J}$,
respectively, by $r_+^2+ r_-^2=
8G\ell^2
m$ and $r_+r_-
=
4G\ell {\cal J}$.
To have a black hole one should impose $r_+\geq r_-$, i.e., 
$m\geq \frac{{\cal J}}{\ell}$.
The inequalities are saturated in the extremal case, $r_+=r_-$, 
i.e., $m=\frac{{\cal J}}{\ell}$.

Without introducing a regulator boundary, the on-shell Euclidean
action (\ref{act0}) diverges.
By introducing the regulator boundary at $r=\bar r$,
the on-shell Euclidean action reduces to 
\bea
S_E(\bar r)=\frac{b_+}{4G\ell^2} \big(\bar r^2-r_+^2\big),
\eea
where $b_+$ is the inverse Hawking temperature 
\bea
 b_+=
\frac{2\pi \ell^2 r_+}{r_+^2-r_-^2}\,,
\label{hawkingapp1}
\eea
given by 
the period of the Euclidean time.  The divergent part in the limit of
$\bar r\to\infty$ arises because of the asymptotically AdS structure in
the (2+1)-dimensional spacetime, and hence from $S_E(\bar r)$ we
should subtract the vacuum AdS counterpart
\bea
S_E^{(0)}(\bar r)=\frac{b_0 (\bar r)}{4 G\ell^2}{\bar r}^2\,,
\eea
for some temperature $b_0$.
The periodicity of the vacuum  AdS is chosen by requiring that the
periodicity of the Euclidean time and the geometry at the section of
$r=\bar r$ in the BTZ and AdS backgrounds should be identical, namely,
\bea
\label{prd}
b_0(\bar r)f_0(\bar r)=b_+ f(\bar r) \,,
\eea
where $f(r)$ is given in Eq.~(\ref{fr}) and 
$f_0(r)$ is given by the vacuum case, i.e.,  
\bea
f_0(r) =\frac{r}{\ell}.
\eea
With \eqref{prd},
defining the regularized Euclidean action by 
\bea
S_{\rm reg}(\bar r)= S_E(\bar r)-S_E^{(0)}(\bar r)
=\frac{b_+}{4G\ell^2}
\Big(
{\bar r}^2-r_+^2
-\frac{b_0(\bar r)}{b_+}{\bar r}^2
\Big)
=\frac{b_+}{4G\ell^2}
\Big(
{\bar r}^2-r_+^2
-\frac{f(\bar r)}{f_0(\bar r)}{\bar r}^2
\Big),
\eea
and expanding it in terms of the inverse power of ${\bar r}$,
we find that
the divergent terms of $O(\bar r^2)$ are canceled out,
and in the limit of $\bar r\to \infty$
\bea
S_{\rm reg}=-\frac{\pi r_+}{4 G}.
\eea
The Euclidean action $S_{\rm reg}$ is related to the free energy $F$
by $S_{\rm reg} = b_+ F$, hence
\bea
F=
\frac{S_{\rm reg}}{b_+}
=-\frac{r_+^2-r_-^2}{8 G\ell^2}\,.
\eea
Now, the first law of black hole thermodynamics is 
\bea
dE=T_+dS+\Omega_+ d{\cal J}\,,
\label{1stlaw1}
\eea
where $E$ is the spacetime energy, 
$T_+$ the black hole temperature 
(given by the inverse of Eq.~(\ref{hawkingapp1}), 
$T_+=1/b_+$), $S$ the entropy, 
$\Omega_+=\frac{r_-}{\ell r_+}$ the 
black hole angular velocity,
and ${\cal J}$ the spacetime angular momentum.
With the definition of the free energy as
\bea
F= E - T_+ S -{\cal J}\Omega_+ \,,
\label{free1}
\eea
the first law of the black hole thermodynamics (\ref{1stlaw1})
can be rewritten as 
\bea
dF=-S dT_+ - {\cal J} d\Omega_+.
\eea
Thus, the entropy and angular momentum of the black hole,
conjugate to $T_+$ and $\Omega_+$ respectively,
are obtained as
\bea
\label{ent_am}
S
&=&
-\Big(\frac{\partial F}{\partial T_+}\Big)_{\Omega_+}
=\frac{\ell^2\pi^2 T_+}{G(1-\ell^2\Omega_+^2)}
=
\frac{\pi r_+}{2G}
=\frac{A_+}{4G},
\nonumber\\
{\cal J}&=&
-\Big(\frac{\partial F}{\partial \Omega_+}\Big)_{T_+}
=\frac{\ell^4 \pi^2 T_+^2\Omega_+}{G(1-\ell^2\Omega_+^2)^2}
=\frac{r_+ r_-}{4 G\ell},
\eea
where $A_+=2\pi r_+$ is the area --- or, in this context, the circumference ---
of the event horizon.
The entropy $S$
is the Bekenstein-Hawking entropy. 
The second relation confirms the expression
for the spacetime angular momentum ${\cal J}$. Finally, the energy $E$
is given by
\bea
E=F + T_+S +{\cal J}\Omega_+
=\frac{r_+^2+r_-^2}{8 G \ell^2}=m\,,
\eea
which means that the thermodynamic 
energy stored in the gravitational system
is given by the ADM mass $m$.

%%%%%%%%%%%%%%%%%%%%%%%%%%%%%%%%%%%%%%%%%%%%%%%

\end{document}